\documentclass[aps,pra,superscriptaddress,,twocolumn,showpacs]{revtex4}
\usepackage{amsmath}
\usepackage{graphicx}
\usepackage{amssymb}
\usepackage{bbm, color}

\newcommand{\ket}[1]{\left|#1\right>}

\newcommand*{\1}{{\mathbbm{1}}}
\newcommand{\T}{\mbox{$\textsf{tr}$}}

\begin{document}

\title{Secure sequential transmission of quantum information}

\author{Kabgyun Jeong}
\author{Jaewan Kim}
\affiliation{School of Computational Sciences, Korea Institute for Advanced Study,
  Hoegiro 85, Dongdaemun, Seoul 130-722, Korea}

\date{\today}
\pacs{03.67.-a, 03.67.Dd, 03.67.Hk}

\begin{abstract}
We propose a quantum communication protocol that can be used to transmit any quantum state, one party to another via several intermediate nodes, securely on quantum communication network. The scheme makes use of the sequentially chained and approximate version of private quantum channels satisfying certain commutation relation of $n$-qubit Pauli operations. In this paper, we study the sequential structure, security analysis, and efficiency of the quantum sequential transmission~(QST) protocol in depth.
\end{abstract}

\maketitle

\section{Introduction} \label{intro}

One of the most popular quantum cryptographic primitives, except quantum key distribution, is the quantum secret sharing~(QSS) protocols~\cite{HBB99,CGL99}. The primitive known as QSS is a process of splitting a piece of quantum information into several parts, and then securely reconstructing the information, but certain subparts are not enough to restore the original quantum information. (In a strict sense, the secret sharing is different from the state sharing on its goal~\cite{KKI99}, but we treat the protocols as in the same category.) There are huge number of theoretical studies on QSS protocols, and also exist experimental demonstrations on QSS schemes in continuous-variable regime, e.g., Ref.~\cite{TZG01,LSBSL04}.

Transmission or distribution of quantum information over several authorized nodes is essential for future applications in quantum communications. We here review the original QSS scheme from the point of view of (approximate) private quantum channels (PQC), and then propose an information transmission method, namely, $\varepsilon$-secure quantum sequential transmission (QST) protocol. This protocol uses a concept of private quantum channel and aims to secure sequential transmission, where arbitrary quantum states pass through several authorized intermediate nodes (or participants). In the transmission process, all nodes must collaborate to reveal the original quantum information. In the sequentially chained scheme, we exploit the Pauli commutation relations on $n$-qubit quantum states, and derive the mathematical structure of multi-node $\varepsilon$-randomizing maps.

Using the idea of the general three-party QSS scheme, we construct our main protocol known as quantum sequential transmission protocol under the security parameter $\varepsilon$. The parameter $\varepsilon$ implies that security and efficiency of the protocols are dealt with an asymptotic consideration. Shortly speaking, the quantum sequential transmission protocol can transmit any quantum states from one party to another under the consent of all authorized participants with classical secret bits. Note that any input quantum states into a quantum channel are arbitrary quantum information with a given dimension, and we exclude classical information. Thus we expect that the protocol, QST, can be applied to certain applications such as quantum repeater~\cite{SSRG11}, quantum key repeater~\cite{BCHW14}, quantum sealed-bid auction~\cite{N09} or quantum email protocol, since only authorized users can send, confirm, and read the quantum message. Furthermore, with the proposed schemes, we study the key question of finding minimal resources required to split and reconstruct a quantum state, and to transfer arbitrary quantum information sequentially.

Let us briefly review the quantum one-time pad or private quantum channel~(PQC). Ambainis \emph{et al}.~\cite{AMTW00} first proposed a quantum primitive known as a private quantum channel for secure transmission of quantum states, and already proved its security including the optimality~\cite{NK06,BZ07}. The complete randomization method naturally gave births to \emph{approximate} approaches for randomizing quantum states~\cite{HLSW04,AS04,DN06}. We here adopt an approximate version of the Dickinson and Nayak's PQC~\cite{DN06}, which has relatively few Pauli operations on multi-qubit encodings. Using conventions and definitions in Sec.~\ref{background}, we construct a quantum communication protocol that are efficient and secure with a small information leakages~($\varepsilon\ll1$) notwithstanding minimal use of resources. But, in this paper, we mainly focus our attention on constructing the mathematical structure of the $\varepsilon$-secure quantum sequential transmission scheme (QST).

Before finishing the section, we introduce the basic concept of (approximate) PQC or $\varepsilon$-randomizing map (or also known as random unitary channel). Moreover we comment on security analysis from Holevo bound and the correspondence between (classical) keys and Pauli operations. 
In Section~\ref{QST}, we focus on our main construction of quantum sequential transmission protocol, which is one step more advanced form of the well-known three-party QSS, on multi-party system. In Section~\ref{conclusion}, we summarize and conclude our work.

\subsection{Background}\label{background}
For a given $d$-dimensional Hilbert space ${\mathbb{C}^d}$, $\mathcal{B}({\mathbb{C}^d})$ denotes
the space of bounded linear operators on the space ${\mathbb{C}^d}$, and $U(d)\subset\mathcal{B}({\mathbb{C}^d})$ the unitary group on the space. We make use of a quantum map from $\mathcal{B}({\mathbb{C}^d})$ to itself generally known as a quantum channel. Then we can define a private quantum channel: For any $\varepsilon\ge0$, a completely positive and trace preserving~(CPTP) map
$\mathcal{R}:\mathcal{B}({\mathbb{C}^d})\to\mathcal{B}({\mathbb{C}^d})$ is said to be $\varepsilon$-randomizing with respect to the trace norm if, for all quantum state $\rho\in\mathcal{B}({\mathbb{C}^d})$,
\begin{equation}
\left\|\mathcal{R}(\rho)-\frac{\1_d}{d}\right\|_1\le\varepsilon,
\label{epsran}
\end{equation}
where $\1_d$ denotes the identity matrix of a given dimension $d$. The input quantum source $\rho$ is a $d$-dimensional density matrix. The map $\mathcal{R}$ satisfying Eq.~(\ref{epsran}) is the approximate private quantum channel~(APQC), and $\|M\|_1:=\T\sqrt{M^{\dagger}M}$ denotes the trace norm for any matrix $M$. Note that the mapping $\mathcal{R}$ is perfect (or complete) randomizing map if $\varepsilon=0$. The definition with the security parameter $\varepsilon$ always implies perfect PQC, and gives us the informational security rather than a security based on the computational complexity.

A simple way to create such an invertible encoding map is to choose a certain sequence of unitary operators $U_1,\ldots,U_{s\le d^2}\in U(d)$ and define the encoding map as
\begin{equation}
\mathcal{R}:\rho\to\frac{1}{s}\sum_{i=1}^s U_i\rho U_i^{\dagger}.
\label{RUC}
\end{equation}
The index $i$ corresponds to the number of shared secret bits that all communicating parties share. We here assume that the secret bits are unknown to any eavesdroppers or unauthorized parties.
With a suitable choice of $s$ unitary operators not more than $d^2$, the mapping $\mathcal{R}$ satisfies to be an approximate private quantum channel. In fact, any orthogonal set of $d^2$ unitary operations form a perfect private quantum channel. Notice that the dimension $d$ of our case is fixed
to $2^n$ to accommodate the Hilbert space of $n$ qubits.

If that is the case, how can we analyze the security of approximate private quantum channels? Roughly speaking, the accessible information to any attackers, for any quantum states $\rho=\sum_ip_i\rho_i$ supported on $\mathbb{C}^d$ and $d\varepsilon<1$, is bounded above by Holevo information~\cite{H73}
\begin{eqnarray*}
\chi\{p_i,\mathcal{R}(\rho_i)\}
&=&S\left(\sum_{i=1}^dp_i\mathcal{R}(\rho_i)\right)
-\sum_{i=1}^dp_iS(\mathcal{R}(\rho_i)) \\
&\le&\log(1+d\varepsilon)<d\varepsilon,
\label{Holevo}
\end{eqnarray*}
where $\{p_i,\mathcal{R}(\rho_i)\}$ represents an ensemble of $\rho_i$'s with probability $p_i$'s through the quantum channel $\mathcal{R}$, and $S(\rho):=-\T\rho\log\rho$, the von Neumann entropy.
The above inequality is true because the definition of the $\varepsilon$-randomizing map with respect to the trace norm in Eq.~(\ref{epsran}) implies that the eigenvalues of the channel-output are almost uniformly distributed such that $\mathcal{R}(\rho_i)\simeq (1+d\varepsilon)\1_d/d$. This also means that attackers cannot obtain any information about the information of the ensemble $\{p_i,\mathcal{R}(\rho_i)\}$ under the condition $d\varepsilon<1$.

Finally, a relation between keys and Pauli operators is crucial in the proof of following protocol, so we carefully investigate the key correspondence. An explicit construction for Eq.~(\ref{RUC}) depends on unitary operators chosen at random from the set of $n$-qubit Pauli matrices. For two $n$-bit strings $a$ and $b$, let $a*b=\sum_{j=1}^{n}a_j b_j\mod 2$ denotes the standard inner product on $\mathbb{Z}_2^n$. We represent a tensor product of $n$ single qubit Pauli operators by a string of $2n$-bit $K, (a,b)\in \{0,1\}^{2n}$,
by using the following correspondence
\begin{equation}
K=(a,b)~~~~:~~~~\iota^{a* b}X^{a}Z^{b},
\label{nPauli}
\end{equation}
where $X^{a}Z^{b}=X^{a_1}Z^{b_1}\otimes\cdots \otimes X^{a_n}Z^{b_n}$ with $X=\begin{pmatrix}
    0  & 1 \\
    1  & 0 \\
    \end{pmatrix}$ and $Z=\begin{pmatrix}
    1  & 0 \\
    0  & -1 \\
    \end{pmatrix}$,
and $\iota=\sqrt{-1}$ the imaginary number. Now, we define a set $P_n$ as $$\left\{\iota^{a* b}X^{a}Z^{b}:(a,b)\in \{0,1\}^{2n}\right\}\subset U(2^{n})$$ for all tensor products of $n$ single qubit Pauli operators. Then the set $P_n$ forms a basis for the $2^{n}\times2^{n}$ complex matrices. (Note that the set $P_1=\{\1_2,X,\iota XZ,Z\}$ is the usual Pauli operators on single qubit.) For convenience we substitute $P_n$ to $P_K$ under the correspondence in Eq.~(\ref{nPauli}) to emphasize a classical key $K$.

As we mentioned above, $n$-qubit Pauli operators form a basis for the set of all $2^n\times2^n$ matrices. So, for a given density matrix $\rho$, we can construct that
\begin{equation}
\rho=\frac{1}{2^n}\sum_{(a,b)\in\{0,1\}^{2n}}c_{a,b}X^aZ^b,
\label{paulidecom}
\end{equation}
where $c_{a,b}$ is an element of a vector $(c_{a,b})$ in $\mathbb{C}^{2^{n}}$ with $\|c_{a,b}\|_2^2\le2^n$, and $\|X\|_2:=\sqrt{\T X^\dagger X}$ is the Frobenius (or Hilbert-Schmidt) norm on the space.

\section{Quantum sequential transmission scheme}\label{QST}
With additional modification of QSS~\cite{HBB99,CJ14,SGYLW12} and approximate private quantum channels,
we now propose a quantum transmission protocol of so-called $\varepsilon$-\emph{secure} quantum sequential transmission (QST) scheme. The main objective of our task is to send a unknown quantum information from a sender to a receiver when several authorized intermediate nodes exist. Although the quantum information is transmitted sequential ways on concatenated quantum channels, the crucial advantage of this protocol is to preserving its explicit security and efficiency. In the sequential structure, we take the generalized $n$-qubit Pauli commutation relations on any input quantum signal, and prove the mathematical consistency and security of the chained $\varepsilon$-randomizing maps. 

As in three-party QSS protocol, suppose that, for all $i$-th position, Alice, Bob and Charlie share a correlation key such that $k_i^A\oplus k_i^B\oplus k_i^C=\alpha_i\mod2$, where $\alpha_i$ is fixed to 0 under the mod 2 operation. The main purpose of this protocol is to securely transmit a quantum state from Alice to Charlie through a middle party Bob. The transmitted state between Alice and Charlie is asymptotically secure since the $2n$-bit-key-based PQC makes arbitrary $n$-qubit state into a \emph{near} maximally mixed state (in three-party scenario). Extending the idea of three-party protocol, we can directly generalize it to an $m$-party concatenated-transmission protocol within $n$-qubit Pauli commutation relations.

First, we simply take account of three-party protocol for sequential quantum state transmission. Alice prepares an $n$-qubit quantum state $\ket{\Phi}\in\mathcal{B}(\mathbb{C}^{2^{n}})$ and encodes the state to $P_{K^A}\ket{\Phi}$ which will be transmitted to Bob. (Consideration of only pure states is enough since the convexity of trace norm ensures the previous statement.) Bob also encodes the state, by using the correlation key $K^B$, to $P_{K^B}\circ P_{K^A}\ket{\Phi}$ where $\circ$ denotes a composition of two Pauli sets, and sends the state to the third party Charlie. Remember that $k_i^A\oplus k_i^B\oplus k_i^C=\alpha_i$, so the receiver Charlie efficiently decodes the state to original quantum information
\begin{equation}
P_{K^C}\circ P_{K^B}\circ P_{K^A}\ket{\Phi}=\ket{\Phi}.
\label{decode}
\end{equation}
In Eq.~(\ref{decode}), we use the following identity,
for any Pauli operators,
\begin{equation}
P_{K^C}\circ P_{K^B}\circ P_{K^A}=P_{K^C\oplus K^B\oplus K^A\mod 2}=P_0:=\1_{2^n}.
\label{commute}
\end{equation}
This condition for (complete) private quantum channel needs exactly $2n$ secret bits. But if we use the approximate case of PQC, then we need  about half-size ($\approx n$-bit) keys only instead of $2n$-bit keys \cite{HLSW04,DN06}, i.e., for sufficiently large $d$ it satisfies our security level with small $\varepsilon$.

\begin{figure}
  \includegraphics[angle=0,width=0.95\linewidth]{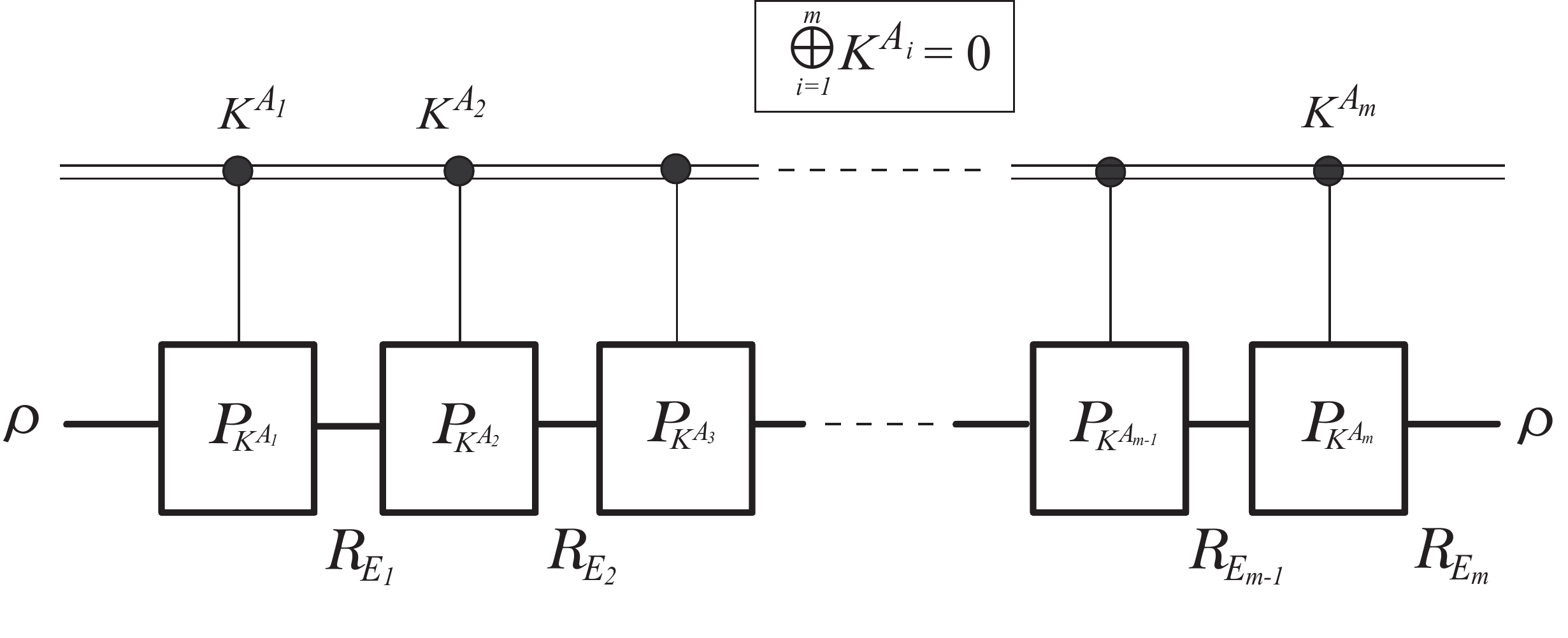}
\caption{Approximate $m$-party quantum sequential transmission protocol: By using a secret classical information $K$, a sender transmits any quantum state $\rho$ securely to final node through the $m-1$ and $\varepsilon$-randomizing maps $\mathcal{R}_{E_j}$ for all $j$. Boxes with $P_K$ represent the $n$-qubit Pauli operations corresponding a key $K$.}
\label{Fig:QST2}
\end{figure}

As shown in Fig.~\ref{Fig:QST2}, $m$-party extension~($m\ge3$) of QST scheme is simple and natural, but we need some technical calculations as shown below. We note that every intermediate user also accomplishes the role of sender and receiver. Before describing the $m$-party scenario, we define a bias of a set of $2n$-bit strings. For a given subset of $E\subset\{0,1\}^{2n}$, if
\begin{equation}
Bias(E,(a,b))=\left|\mathbb{E}_{x\in E}(-1)^{x*(a,b)}\right|,
\label{bias}
\end{equation}
then we call the set $E$ is \emph{biased} with respect to a string $(a,b)\in\{0,1\}^{2n}$~\cite{DN06,NN93}, where $\mathbb{E}$ is an expectation value for some variable in $E$. Note that $*$ is also the inner product, and the bias is equal to $2\mathbb{E}_E~[x*(a,b)]-1$ under the modulo 2 operations.
When, for all $(a,b)\neq0^{2n}$, $Bias(E,(a,b))\le\beta$, we call the subset $E\subset\{0,1\}^{2n}$ to be $\beta$-biased.

A subset $E\subset\{0,1\}^{2n}$ defines a CPTP map on $n$-qubit as follows,
\begin{eqnarray}\label{RErho}
\mathcal{R}_E(\rho)&=&\frac{1}{|E|}\sum_{(u,v)\in E}X^uZ^v\rho Z^vX_u \nonumber\\
&=&\frac{1}{2^n|E|}\sum_{(u,v),(a,b)}c_{a,b}
X^uZ^v(X^aZ^b)Z^vX^u \nonumber\\
&=&\frac{1}{2^n}\sum_{(a,b)\in\{0,1\}^{2n}}c_{a,b}\beta_{a,b}X^aZ^b,
\label{nrchannel}
\end{eqnarray}
where a real number $|\beta_{a,b}|$ is equal to the $Bias(E,(a,b))$ in Eq.~(\ref{bias}). The modulus of $E$, $|E|$, corresponds to some number $s(\le 2^{2n})$ of $n$-qubit Pauli operations used in the
map $\mathcal{R}_E$. By using commutation relations on Pauli matrices, above equations
can be derived from
\begin{eqnarray}\label{XY}
X^uZ^v(X^aZ^b)Z^vX^u
&=&(-1)^{a* v+b* u}X^aZ^b \\
&=:&\beta_{a,b}X^aZ^b. \nonumber
\end{eqnarray}
If we choose $E=\{0,1\}^{2n}$, then we have a completely randomizing map. It is known that there \emph{exists} a map $\mathcal{R}_E$ an $\varepsilon$-randomizing map with respect to the trace norm for $n$-qubit states, when the subset $E\subset\{0,1\}^{2n}$ be a set with bias at most $\varepsilon\cdot2^{-n/2}$. (See also the proof in Ref.~\cite{AS04}.)

From the existence of small $\beta$-bised subset $E$, the Frobenius norm of the randomized state is almost concentrated at the maximally mixed state, that is,
\begin{equation}
\label{Fbound}
\|\mathcal{R}_E(\rho)\|_2^2\le\frac{1+\varepsilon^2}{2^n}.
\end{equation}
This inequality can be directly calculated from the Eq.~(\ref{RErho}) of $\varepsilon\cdot2^{-n/2}$-biasedness and the bound $\|c_{a,b}\|_2^2\le2^n$.

Moreover, for any density matrix $N\in\mathcal{B}(\mathbb{C}^{2^n})$, the inequalities $\|N\|_1\le\sqrt{2^n}\cdot\|N\|_2$ and $\left\|N-\frac{\1_{2^n}}{2^n}\right\|_1^2\le2^n\cdot\|N\|_2^2-1$
always hold. (See proof details in the appendix A of Ref.~\cite{DN06}.) Thus we obtain the following chain bounds
\begin{equation}
\label{epsran2}
\left\|\mathcal{R}_E(\rho)-\frac{\1_{2^n}}{2^n}\right\|_1\le
\sqrt{2^n\|\mathcal{R}_E(\rho)\|_2^2-1}\le
\varepsilon.
\end{equation}
Thus, if we can choose a suitable subset $E$ with $\beta$-biasedness, then we can always create $\varepsilon$-randomizing map or APQC in trace norm. The above equation, Eq.~(\ref{epsran2}), is intrinsically identical to the Eq.~(\ref{epsran}), therefore the security is well preserved.

Finally we show that multi-party approximate private quantum channel and multi-party quantum sequential transmission protocol is secure and efficient, i.e., we claim that $n_{\mathnormal{DN}}:=n+2 \log{\frac{1}{\varepsilon}}+4$ classical keys are sufficient for the $m$-party QST scheme. By choosing a dense subset $E$, we can initialize a subset $E_j\subset\{0,1\}^{2n}$ to be a set with bias
at most $\varepsilon^{1/m}\cdot2^{-n/2m}$ for each $j$~\cite{J13}. Then we assert that there exists an $m$-party $\varepsilon$-randomizing map with respect to the trace norm for $n$-qubit states: For any density matrix $\rho\in\mathcal{B}(\mathbb{C}^{2^n})$, we have
\begin{equation}
\left\|\mathcal{R}_{E_m}\circ\mathcal{R}_{E_{m-1}}\circ\cdots
\circ\mathcal{R}_{E_1}(\rho)-\frac{\1_{2^n}}{2^n}\right\|_1
\le\varepsilon.
\end{equation}
We here denote that $\mathcal{R}_T=\mathcal{R}_{E_m}\circ\cdots\circ\mathcal{R}_{E_1}$ for convenience. Since the $m$-user encoding and transmitting for a quantum state under $m$-APQC form an $m$-party QST protocol, and it can be directly derived from the following commutation relation
\begin{widetext}
\begin{equation}
X^{u_m}Z^{v_m}\cdots (X^{u_1}Z^{v_1}(X^aZ^b)Z^{v_1}X^{u_1})\cdots Z^{v_m}X^{u_m}
=(-1)^{\sum_{j=1}^m a* v_j+b* u_j}X^aZ^b.
\end{equation}
\end{widetext}
This equation is just a generalization of Eq.~(\ref{XY}). Suppose that, for every quantum state $\rho\in\mathcal{B}(\mathbb{C}^{2^n})$, each $\varepsilon$-randomizing map between two nodes $(j,j+1)$ satisfies
\begin{equation}
\left\|\mathcal{R}_j(\rho)-\frac{\1_{2^n}}{2^n}\right\|_1\le\varepsilon^{\frac{1}{m}},
\label{meps}
\end{equation}
then we can always construct multi-user QST protocol via approximate private quantum channels with
\begin{equation}
\left\|\mathcal{R}_T(\rho)-\frac{\1_{2^n}}{2^n}\right\|_1\le
\varepsilon,
\label{teps}
\end{equation}
and consume about $n$ bits of secret classical keys satisfying $\bigoplus_{i=1}^m {K^{A_i}}=0$. This result implies that QST based on sequential private quantum channels is secure. The estimation of Eq.~({\ref{teps}}) for every $\varepsilon$ promises to use the classical key of $n+2 \log{\frac{1}{\varepsilon}}+4$ bits~\cite{DN06}. Notice that Dickinson and Nayak's efficient construction for the approximate PQC on $n$-qubit situation relies on McDiarmid's inequality in probability analysis and a net argument on discretizing pure quantum states. Strict security analysis for the approximate private quantum channel in security parameter $\varepsilon$ are reported at Ref.~\cite{BZ07}, and see also Ref.~\cite{ZB05}.

\section{Conclusion}\label{conclusion}
In summary, we constructed a quantum communication protocol for quantum sequential and $\varepsilon$-secure transmission scheme via the extension of three-party QSS task. This scheme makes use of a relatively small (correlated) classical secret information of about $n_{\mathnormal{DN}}\simeq n$ bits, just half of the size or the perfect private  quantum channel of $2n$-bit, and transmit any $n$-qubit states securely, so we say that the protocol is efficient. The security argument only depends on the small security parameter $\varepsilon$ in which an approximate private quantum channel guarantee its security. In fact, it is a small value ($\varepsilon< 1$) for sufficiently large $d$-dimension of Hilbert space $\mathbb{C}^d$.

Beyond the mathematical construction of the quantum sequential transmission scheme, we need to exploit this type of communication protocols for potential future applications such as quantum (key) repeater, auction, and email scheme and so on. So, the analysis of these protocols in quantum regime is significant and necessary. We finally point out that the security of the QST protocol must be systematically analyzed for several cases of attackers, and further study is needed for mathematical generalization in $p$-norm cases (for all $p>1$). We hope that the quantum sequential transmission, QST, can be used for the realization of practical quantum communication networks.

\section{acknowledgement}
This work was supported by Korea Institute for Advanced Study~(KIAS) grant funded by the Korea government (MSIP) and partly supported by the IT R\&D program of MOTIE/KEIT [10043464].


\begin{thebibliography}{26}
\bibitem{J13}
There exixts such a subset $E_j\subset\{0,1\}^{2n}~\forall j$ with
$\sqrt[m]{\varepsilon}/2^{n/2m}$-biased, because the subset $E$ in
Eq.~(\ref{bias}) is upper bounded by $\varepsilon/2^{n/2}$.
\end{thebibliography}
\end{document}